\documentclass[showpacs,aps]{revtex4}
\usepackage{graphicx}

\begin{document}

\title{A three-dimensional scalar field theory model of center vortices and its relation to ${\bf k}$-string tensions}
\author{John M. Cornwall\footnote{Email:  Cornwall@physics.ucla.edu}}
\affiliation{Department of Physics and Astronomy, University of California, Los Angeles CA 90095}

\begin{abstract}
\pacs{11.15.-q, 12.38.-t, 11.15.Tk   \hfill UCLA/04/TEP/5}
In $d=3$ $SU(N)$ gauge theory, we study a scalar field theory model of center vortices, and their monopole-like companions called nexuses, that  furnishes an approach to the determination of so-called $k$-string tensions.  This model is constructed from string-like quantum solitons introduced previously, and exploits the well-known relation between string partition functions and scalar field theories in $d=3$.  A basic feature of the model is that center vortices corresponding to magnetic flux $J$ (in units of $2\pi /N$) are composites of $J$ elementary $J=1$ constituent vortices that come in $N-1$ types, with repulsion between like constituents  and attraction between unlike constituents.   The scalar field theory is of a somewhat unusual type, involving $N$ scalar fields $\phi_i$ (one of which is eliminated) that can merge, dissociate, and recombine while conserving flux $mod\;N$.  The properties of these  fields are deduced directly from the corresponding gauge-theory quantum solitons.     Every vacuum Feynman graph of the theory corresponds to a real-space configuration of center vortices. We use qualitative features of this theory based on the vortex action to study the problem of $k$-string tensions (explicitly at large $N$, although large $N$ is in no way a restriction on the model in general), whose solution is far from obvious in center-vortex language.  We construct a simplified dynamical picture of constituent-vortex merging, dissociation, and recombination, which allows in principle for the determination of vortex areal densities and $k$-string tensions.    This picture involves point-like ``molecules" made of constituent ``atoms" in $d=2$ which combine and disassociate dynamically.  These molecules and atoms are cross-sections of vortices piercing a test plane; the vortices evolve in a Euclidean ``time" which is the location of the test plane along an axis perpendicular to the plane.  A simple approximation to the molecular dynamics is compatible with  $k$-string tensions that are linear in $k$ for $ k\ll N$, as naively expected.  
\end{abstract}

\maketitle

\section{Introduction}
\subsection{General remarks}

Center vortices have been intensively studied since their introduction more than twenty-five years ago \cite{thooft78,corn79,mackpet,acy,nielsole,etomb}.
There are now many lattice studies (for a modern review, see \cite{green04}), which confirm the basic topological mechanism for confinement in center-vortex theory. 

In this paper we present a model for center vortices and nexuses in $d=3$, based on the well-known connection \cite{stonet} between closed-string partition functions and (scalar) field theories in three dimensions.  This model is applicable, in principle, for all $SU(N)$ gauge theories.  In $d=4$ there is an analogous model, not explored here, which would be a closed-string theory of unusual type.  In both $d=3$ and $d=4$ we would find exactly the same picture described below of ``atoms" and ``molecules" interacting in a plane, as a description of center-vortex areal densities.  These densities are essential to our understanding of confinement.

While it is desirable in general to have a good model for center vortices that  is simpler than directly solving the full gauge theory, the real point of the model is the extent to which it can clarify the full theory.  We have in mind here applying the  model to the question of string tensions with non-trivial $N$-ality, which has presented some difficulties in center-vortex theory.  This remains one of the  outstanding first-principle problems of confining gauge theories in the center vortex picture.  

The  primary focus of the present paper is the model itself, but before presenting the model we will make some remarks on the problem of   the so-called $k$-string tension, which is the string tension $\sigma (k)$ that arises between a test particle (in $SU(N)$ gauge theory) whose group representation is described by a Young tableau of a single column of $k<N$ boxes, and its antiparticle, described by $N-k$ boxes.  This test particle and its antiparticle form a singlet and are confined.  The $k$-string problem exists only for $N\geq 4$, since for $N=2,3$ the $k$ representations consist only of the fundamental and antifundamental representations.  But for larger $N$ one expects to have different string tensions for different $k$.  

Various simple arguments, for example, standard large-$N$ considerations, suggest (at least if $k\ll N$) that $\sigma (k)\sim k$.  Or one could say that a $k$ representation is the totally antisymmetric product of $k$ fundamental representations (quarks), so that in terms of the mesonic string tension  $\sigma (1)$ the $k$-string tension ought to be composed of $k$ quark-antiquark strings and hence $k$ times as big.  

Unfortunately, it has not proven to be so easy to derive $\sigma (k)\sim k$ in the center vortex picture of confinement.  The reason is that in the center vortex picture confinement arises from the topological linking of center vortices with Wilson loops, and it is not easy to envisage the $k$ separate quark-antiquark strings mentioned above.  In the center vortex picture,  a quantity from which $\sigma (k)$ can be found is the discrete Fourier transform of the areal (two-dimensional) density of center vortices with different values $L$ of magnetic flux, measured in units of $2\pi /N$, and one needs to find these areal densities as a function of $L$.

With no compelling picture of vortex areal densities, the result so far has been that theoretical approaches to $\sigma (k)$ in the center vortex picture have been largely phenomenological.  Workers often take Casimir scaling of $\sigma (k)$ or analogs to M-theory as the two possible standard forms for the $k$-string tensions.  These forms are
\begin{equation}
\label{kforms}
\sigma_{Cas}(k)=\sigma_1\frac{k(N-k)}{N-1},\;\sigma_{Mth}(k)=\sigma_1\frac{\sin (\pi k/N)}{\sin (\pi /N)}.
\end{equation}
Both of these lead to $\sigma (k)\sim k$ for  $k\ll N$, but they have different terms at non-leading order  for large $N$.  There are arguments \cite{arshif} that non-leading terms exclude Casimir scaling, but in fact which of these two forms, or some other form, actually holds is not known now.  In any case, we will not study non-leading terms here.  Both forms above also obey the necessary group-theoretic conjugation symmetry $\sigma (k)=\sigma (N-k)$.  
Certain works that address $\sigma( k)$ in the center vortex picture depend on some form of quadratic approximation that inevitably leads to Casimir scaling \cite{antdel}; others are phenomenological, and ask what density of vortices is needed to produce $\sigma (k)$ as given by the above choices \cite{deldia,go} or other related forms.  Using the dilute gas approximation (DGA), Ref. \cite{deldia} finds unphysical areal densities amounting to the unchecked pileup of center vortices at a single point.  Ref. \cite{go} (hereafter GO) supplies a relation between $k$-string tension and areal densities which is usable in principle where the DGA is not, and we will make use of this relation in the present paper.

In addition to these theoretical works, there are lattice simulations \cite{luctep,dprv,luctep3,dpv} for $N=4,5,6$.  Most of this lattice work supports the M-theory form of the $k$-string tension, but it does not seem possible to draw unassailable conclusions from the present data.  A very recent work \cite{luctep4} concludes that lattice data lie somewhere between M-theory and Casimir scaling.

So far (to the author's knowledge) there has been no discussion from the fundamental principles of center-vortex theory of the areal densities needed to find the $k$-string tension.  While we do not claim to solve the $k$-string tension problem here, we do offer a path toward solution in the present paper. 

\subsection{Outline of the paper}

The main purpose of the paper is to present the $d=3$ scalar field theory that models the gauge theory.  [There are other models of center vortices; see \cite{er} for $SU(2)$ and \cite {eqr} for $SU(3)$.  But there is no $k$-string problem for these gauge theories.]  A secondary purpose is to introduce the possibility of exploiting the scalar field theory to discover something about $k$-string tensions, although we do not go very far in this direction, and restrict our considerations to large $N$.  [It is not yet known whether center vortices become unimportant at large $N$, although it is very possible \cite{corn98} that they survive in this limit.]   After a quick review of the gauge-theory form of center vortices in sec.~\ref{gaugeth}, we show how to express these gauge-theoretic results via a scalar field theory in sec.~\ref{scalarfth}.   This field theory is constructed directly from the gauge-theory picture of the quantum solitons constituting center vortices and nexuses \cite{corn98}. The basic principles of equating a theory of closed strings, such as center vortices, to a scalar field theory have long been known \cite{stonet}, but the theory we find is considerably more complex than envisaged in these pioneering references.  The scalar theory transcribes the properties of vortices (strings in $d=3$) and nexuses (points in $d=3$) into Euclidean world lines for the particles of the scalar field theory. These strings, with nexuses and antinexuses sitting on them, undergo  processes whose essential features  we will incorporate into the scalar field theory. Specifically, every vacuum graph of this scalar field theory is in one-to-one correspondence with a process of physical vortex strings merging, dissociating, and recombining.
Every center vortex of flux  $J>1$ is describable as a composite of $J$ constituents, each of unit flux.

Sec.~\ref{willoop} goes into some kinematical simplifications which are important for our specific implementation of the model in determining $k$-string tensions, but  it is better to outline first some more general features of the relation of the scalar-field theory model to string tensions, as put forth in sec.~\ref{mastersec}. These general relations will survive the simplifications of sec.~\ref{willoop}.  To describe string tensions, which involve the piercing of $d=2$ surfaces (Wilson loops) by center vortices, we introduce certain probabilities $p(L)$, where $p(L)$ is the probability that a center vortex of flux $L$ intersects a large but finite planar surface, which we call the test plane.   These probabilities are converted to areal densities by dividing by a squared correlation length $\lambda^2$, as GO describe.   We construct a continuous series of cross-sections of $d=3$ vortex processes by rigidly translating  the test plane along its normal (say, the $z$ axis).  The cross-section of a vortex in the plane is point-like, and we refer to composite vortices as  {\em molecules} and the unit-flux constituents as {\em atoms}.  Because of the underlying random-walk nature of the vortices, these atoms and molecules appear to diffuse in the test plane as $z$ changes,  from time to time colliding and interacting with each other.   Under the assumption that such processes are ergodic, averaging over $z$ should yield the same results for areal densities as averaging over many gauge configurations.  We construct a a simple master rate equation in $z$, in an attempt to describe  these merging, recombination, and dissociation interactions.  Which processes are allowed and which are forbidden is governed by a set of force and action rules derived directly from the actions of the vortices in the process. The master rate equation yields (in principle, at least) the vortex areal densities as densities of atoms and molecules in the test plane.   

At this point we do not have enough quantitative information about the scalar field theory model (couplings and masses) to justify a thoroughgoing study of the master rate equation.  Instead, we will make a few suggestive remarks, based on some kinematic simplifications which are motivated by large-$N$ considerations.  These simplifications lead to a modified form of the DGA, which is free from the objectionable features of the phenomenological probabilities $p(L)$ found in \cite{deldia} using the standard DGA approach.  

The kinematic issues are set out in sec.~\ref{largensec}, following GO.
These authors formulate the $k$-string question in a way  that is not subject to the low-density restrictions of the DGA.     They go on to  give purely phenomenological areal densities which both fit a linear rise of $\sigma (k)$ and are well-behaved in the large-$N$ limit. They define $p(L)$ as the probability that a square of side $\lambda$ is pierced by a vortex of flux $L$, where $\lambda$ is the correlation length of vortices. [The probability $p(0)$ is the probability that the square has no vortex in it at all.] They then argue that
\begin{equation}
\label{firstgo}
\exp [-\sigma(k)\lambda^2] =\sum_0^{N-1}p(L)\exp [\frac{2\pi ikL}{N}].
\end{equation}
The inverse discrete Fourier transform yields $p(L)$ in terms of $\sigma (k)$ as
\begin{equation}
\label{firstgoinv}
p(L)=\frac{1}{N}\sum_{k=0}^{N-1}\exp [\frac{-2 \pi ikL}{N}]\exp [-\sigma(k)\lambda^2].
\end{equation}
Conjugation symmetry applied to this inverse transform then yields $p(L)=p(N-L)$ for the areal densities, and then Eq. (\ref{firstgo}) yields  $\sigma (k)=\sigma (-k)$.    In fact, as is well-known, a vortex of flux $N-L$ is actually an anti-vortex of flux $-L$, so we should also have $p(L)=p(-L)$.  All these conjugation-symmetry relations will hold in our basic model for center vortices. 

We are interested in large $N$, and in sec.~\ref{largensec} we explore the consequences of replacing discrete variables and sums over these variables by continuous variables and integrals over these variables.  In so doing, we lose track to some extent of conjugation symmetry, but recover enough of it for our purposes in the evenness of $p(L)$ and $\sigma (k)$.  First, note that Eq. (\ref{firstgoinv}) suggests a scaling symmetry, in which $p(L)$ is replaced by a continuous function:
\begin{equation}
\label{firstscale}
p(L,N)\rightarrow \frac{1}{N}\tilde{p}(x),\;x\equiv \frac{L}{N}
\end{equation}
whereby the sum in the fundamental GO equation (\ref{firstgo}) is replaced by an integral from 0 to 1 over $x$.  Conjugation symmetry for $\tilde{p}(x)$ is retained in the form 
\begin{equation}
\label{firstconjp}
\tilde{p}(x)=\tilde{p}(1-x)\equiv \tilde{p}(-x).
\end{equation}
However, at this point we have lost (see sec.~\ref{largensec}) conjugation symmetry for $\sigma (k)$, and subsequent results for this quantity apply only for $k\ll N$.  By the usual rules of Fourier transforms this replacement of a sum by an integral requires us to turn the sum over $k$ of the inverse transform in Eq. (\ref{firstgoinv}) into  a sum with infinitely many terms.  This, as we will see from explicit GO results, leads to dropping non-scaling terms in $\tilde{p}(x)$ which are exponentially-small in $N$ and is harmless.  

Then we go further and replace the infinite sum over $k$ by an integral over $k$.  We will see in sec.~\ref{largensec} that for any given form of $\sigma (k)$ we must modify $\tilde{p}(x)$ to a new function $\hat{p}(x)$, while at the same time changing the integral over $x$ that expresses $\sigma (k)$ by an integral with infinite limits.  The modified function $\hat{p}(x)$ does not obey $\hat{p}(x)=\hat{p}(1-x)$, but is still even in $x$, which is enough of conjugation symmetry for our purposes.  We see by studying GO's explicit results that $\hat{p}(x)$ is a certain limiting form of $\tilde{p}(x)$, valid when the parameter
\begin{equation}
\label{1stepsilon}
\epsilon \equiv \frac{\sigma (1)\lambda^2}{2\pi}
\end{equation}
is small compared to unity.  We show in sec.~\ref{dgasec} that the smallness of $\epsilon$ is a form of the DGA that is useful, since it is compatible with the scaling of $p(L)$ as given in Eq. (\ref{firstscale}).   It turns out  that requiring $\epsilon \ll 1$ is  appropriate  for justifying the use of Fourier integrals rather than sums.  Taking the usual fundamental string tension, and $\lambda^{-1}\simeq M \simeq$ 600 MeV, we find that $\epsilon$ is about 0.08.

Having made the kinematic simplifications alluded to above, we explore crudely the behavior of the scaling density $\hat{p}(x)$ in the limits of small and large $x$.  This behavior is compatible with, but does not necessarily imply, a linearly-growing string tension.  In fact, the behavior at large flux is precisely that found in \cite{deldia} using the DGA, but our result is compatible with scaling [Eq. (\ref{firstscale})].  

Sec.~\ref{sumsec} summarizes the paper, and in addition there are two appendices covering details of center-vortex actions.

 In future works we will make more extensive studies, both numerical and analytic, of the master rate equation that determines the $p(L)$.
But even at this primitive stage of development,  we hope to convince the reader that it is plausible that the center-vortex picture leads to $\sigma (k)\sim k$ for small $k$.  

\section{\label{gaugeth} Description of center vortices and nexuses in the gauge theory}

Throughout this paper we use gauge potentials that are anti-Hermitean matrices constructed from the canonical potentials and multiplied by the gauge coupling $g$. As discussed in the introduction, we write explicit formulas in $d=3$ Euclidean space, but the transcription to $d=4$ is straightforward ; it leads to an unconventional string theory rather than a field theory.

Center vortices are (quantum) solitons of the effective action, and they are essentially Abelian objects that can be superposed.  An elementary center vortex is described by specifying a closed curve and a group matrix.  The group matrices must be chosen so that the magnetic flux $F$, defined through the Wilson loop as
\begin{equation}
\label{fluxeq}
\frac{1}{D_k}Tr_k\exp [\oint_{\Gamma} dz_iA_i] \equiv \exp [ikF],
\end{equation}
is an element of the center.  Here $Tr_k$ indicates a trace in the  $k$ representation, of dimension $D_k$.  (Of course, $k=1$ is the fundamental representation.)  If the Wilson loop $\Gamma$ is linked
exactly once in a positive sense with the vortex curve, then $F$ is the flux associated with the vortex.  It must have the form $F=2\pi J/N$, where $J$ is an integer from 1 to $N$, in which case we speak of a $J$ vortex.

The idea behind our picture \cite{corn79} of center vortices is that infrared slavery requires the generation of a dynamical gluon mass \cite{corn82}.  This, as seen in Schwinger-Dyson equations for the gluons, induces a gauge-invariant mass term that is simply a gauged nonlinear sigma model.  The effective action for the gauge theory is then 
\begin{equation}
\label{massterm}
I_{eff}=\frac{1}{g^2}\int d^3x[-\frac{1}{2}TrG_{ij}^2- M^2Tr(U^{-1}D_iU)^2]
\end{equation}
where $U$ is an $N\times N$ matrix in the fundamental representation, transforming as $U\rightarrow VU$ when $A_i\rightarrow VA_iV^{-1}+V\partial_iV^{-1}$, and $g^2$ is the square of the coupling constant.  This matrix $U$ is to be eliminated by minimizing the action in which it appears.  If the mass $M$ were considered as a fundamental object, surviving unchanged into the ultraviolet, this effective action would not be renormalizable.  In fact, one knows \cite{lavelle}  that the mass is a function of momentum $q$ decreasing at large $q^2$ as
\begin{equation}
\label{mdecrease}
M^2(q)\rightarrow \frac{const \langle G_{ij}^2\rangle}{q^2}.
\end{equation}
Some such decrease is required for the Schwinger-Dyson equations to be solvable.  We will only use the effective action in the infrared, so this complication of the ultraviolet behavior of the mass term is irrelevant for us.  There is one exception: if the mass $M$ is wrongly taken to be constant, the mass term in $I_{eff}$ has a logarithmic divergence.  Of course, this is not real, and we will ignore it.

An elementary unit-flux center vortex, a solution to the equations of motion of the effective action, has a form essentially equivalent to a Nielsen-Olesen vortex:
\begin{equation}
\label{unitcv}
A_i^{[r]}(x)= \frac{2\pi Q_r}{i}\epsilon_{ijk}\partial_j\oint_{\gamma}dz_i[\Delta_M(z-x)-\Delta_0(x-z)].
\end{equation}
Here $Q_r$ is one of a set of $N$ matrices defined below, whose properties are detailed in Appendix A, and $\Delta_{M(0)}$ is the free Euclidean propagator of mass $M(0)$.  Provided that the curve $\gamma$ is closed, the $\Delta_0$ term is pure gauge, and corresponds to the contribution of the $U$ matrix. (If the curve were not closed, long-range monopoles would sit at the ends of the curve.)  The field strength of the vortex decreases exponentially away from the vortex curve $\gamma$.

The matrices $Q_r$ are
\begin{equation}
\label{qmatrix}
Q_r={\rm diag}(\frac{1}{N},\;\frac{1}{N}, \dots \frac{1}{N},\;-1+\frac{1}{N},\;\frac{1}{N},\;\dots)\;\;r=1,2,\dots N
\end{equation}
with the -1 in the $r$th position.  Each of these matrices obeys
\begin{equation}
\label{qmatrix2}
\exp [2\pi iQ_r]=\exp [2\pi i/N]
\end{equation}
and so has unit flux. Since each $Q_r$ can be transformed into any other by an element of the permutation group, one may think of the index $r$ as a label for group collective coordinates of the vortex. These matrices have the property (see Appendix A) that the sum of all $N$ of them vanishes. 

One may immediately generalize this by adding any number of solitons of the form of Eq. (\ref{unitcv}).  In particular, if for a given curve $\gamma$ we use the flux matrix
\begin{equation}
\label{fluxmatrix}
F_{\{J\}}=Q_{i_1}+\dots Q_{i_J} 
\end{equation}
with all $Q$ indices distinct, we find that the flux of $F_{\{J\}}$ is $J$.  We can construct a conjugate vortex with flux matrix $F_{\{N-J\}}$ by adding together the $N-J$ distinct $Q$ matrices with indices different from all those in $F_{\{J\}}$ of Eq. (\ref{fluxmatrix}).  Because the sum of the $Q$'s is zero, it follows that
\begin{equation}
\label{conjugate}
F_{\{J\}}+F_{\{N-J\}}=0,\;F_{\{-J\}}\equiv -F_{\{J\}}=F_{\{N-J\}}.
\end{equation}
So antivortices of flux $-J$ are entirely equivalent to vortices of flux $N-J$, illustrating the $mod\;N$ nature of flux addition.  As before, the matrix indices on $F_{\{J\}}$ can be shuffled by permutations, which yield group collective coordinates (see Appendix A).   A vortex of flux $J$ is identical in form to that of the unit vortex in Eq. (\ref{unitcv}) except that $Q_j$ is replaced by $F_{\{J\}}$.  In what follows we suppress the collective coordinate index $[r]$ in Eq. (\ref{unitcv}) and similar equations. 

Note that whatever the flux $J$ is, vortex strings always have the same thickness of $M^{-1}$.

\begin{figure}
\includegraphics[height=1.8in]{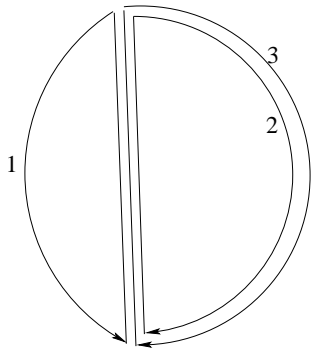}
\caption{\label{merge} Center vortices  merging and dissociating. A $J=1$ vortex (labeled 1 for $Q_1$) meets a $J=2$ vortex ($Q_2+Q_3$) at the bottom vertex to form a $J=3$ vortex, which later dissociates at the top vertex.  }
\end{figure}

It is important for the physical model discussed later to recognize that center vortices can merge and dissociate.  We describe this by a vortex soliton of the form
\begin{equation}
\label{genvortex}
A_i=\frac{2\pi}{i}\epsilon_{ijk}\partial_j\sum_a\oint_a dz(a)_kF_{\{J_a\}}\{\Delta_M[z(a)-x]-\Delta_0[z(a)-x]\}
\end{equation}
where the closed curves labeled by $a$ can have segments in common, as illustrated in Fig. \ref{merge}.

Note that flux is conserved in processes of merging and dissociation as shown in Fig. \ref{merge}.  There is another type of merging allowed, in which flux is only conserved $mod\;N$.  We begin by writing a partial expression for the associated soliton:
\begin{equation}
\label{nsoliton}
A_i(x)=\frac{2\pi}{i}\epsilon_{ijk}\partial_j\sum_{a=1}^N\int_0dz(a)_kQ_a\{\Delta_M[z(a)-x]-\Delta_0[z(a)-x]\}
\end{equation}
where the sum now is over curves $a$ which all begin at the origin.  Because the sum of all $Q_a$ is zero, there are no long-range monopole fields associated with this termination of the curves.  

\begin{figure}
\includegraphics[height=1.8in]{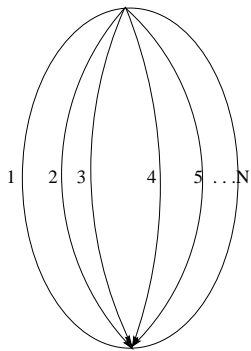}
\caption{\label{baryon} Center vortices of total flux $J=N$ can be annihilated or created from zero flux.  The labels 1, 2, ... correspond to the labels of $Q$ matrices.}
\end{figure}

It is still necessary to close the curves somewhere away from the origin.  One way is simply to repeat the process in Eq. (\ref{nsoliton}) at some other point $x=u$.  This is illustrated graphically in Fig. \ref{baryon}.  Note that this is topologically precisely the same as a bayonic Wilson loop in $SU(N)$.

\begin{figure}
\includegraphics[height=1.8in]{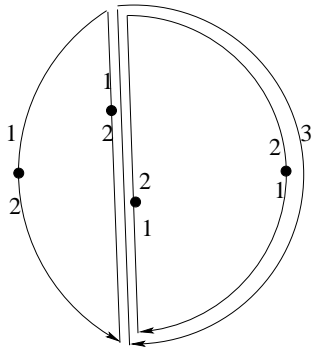}
\caption{\label{recomb} A recombination event, in which vortex 2 (that is, of flux matrix $Q_2$) combines with vortex $\{13\}$ at the bottom, and dissociates into vortex 1 plus vortex $\{23\}$ at the top.  Black circles are nexuses or antinexuses which change 1 into 2 or vice versa.}
\end{figure}

There is one other major type of configuration remaining; it involves nexuses \cite{corn98}.  An elementary nexus is a point-like region on an elementary (unit flux) vortex string where the flux matrix changes smoothly (on the scale set by the mass $M$) from $Q_i$ to $Q_j$.  It is truly a non-Abelian object, with an explicit description too complex for us to record here; it will be enough to know that the nexus is essentially a point on a vortex string which changes the $Q_i$ around.  A nexus (or antinexus) conserves the flux $J\;mod\;N$ but changes the matrices $Q_i$ which make up this flux.  The flux difference $Q_i-Q_j$ is essentially a root vector of $SU(N)$; it has one diagonal element of 1, one element of -1, and the rest are 0. This is the Pauli matrix $\sigma_3$ for some embedded $SU(2)$ of $SU(N)$, and constitutes the flux matrix for what is basically a 't Hooft-Polyakov monopole whose field lines are deformed into the strings of the center vortices on either side of the nexus.  Since every vortex string is closed, every nexus must be accompanied by an antinexus somewhere else on the vortex string.  In $d=4$ the linking of nexus world lines to vortex surfaces leads to topological charge occurring in localized lumps of charge $1/N$, but globally of integral total charge.  This is not of concern to us in three dimensions.  But nexuses do lead to a simple description of important recombination processes, a little different from the simple merging and dissociation of Fig. \ref{merge}.  An example is shown in Fig. \ref{recomb}, showing a $Q_1$ vortex exchanging with a $Q_2$ vortex during merging and dissociation.  
 
\section{ \label{scalarfth} A three-dimensional scalar field model of nexuses and center vortices}

It has long been known \cite{stonet} that the partition function of simple closed strings in $d=3$ can be described by the partition function of a scalar field theory.  The essential ingredient is that the Feynman-Schwinger proper-time representation of the (trace of the logarithm of the) free scalar propagator has an interpretation in the language of closed strings:
\begin{equation}
\label{proptime}
Tr\ln [\Delta_M]=\int_0^{\infty}\frac{ds}{s}\oint(dz)\exp \{-\int_0^sd\tau [\frac{M_1}{2}\dot{z}^2+\frac{M_2}{2}]\}.
\end{equation}
As a stringy object, in the above integral the parameters $s,\tau$ are not proper times but rather physical lengths along the string, and $(dz)$ is an integral over all closed string paths.  The masses $M_{1,2}$ are mass parameters derived from some underlying theory; they are related to the physical mass by $M^2=M_1M_2$. 

For oriented strings, the string partition function is an integral over all closed random walks and lengths:
\begin{equation}
\label{stringpart}
Z_{string}=\sum\frac{1}{N!}{\Big \{} \int_0^{\infty}\frac{ds}{s}\oint(dz)\exp [-\int_0^sd\tau (\frac{M_1}{2}\dot{z}^2+\frac{M_2}{2})]{\Big \}}^N.
\end{equation}
Here the $\dot{z}^2$ term is the familiar action for random walks of length $s$.  It involves a mass factor $M_1$ related to the diffusion rate of the random walks.  The other mass $M_2$ gets positive contributions from string self-energy per unit length and negative contributions from entropy, so it is possible that $M_2$  and $M^2$ are negative, and the field is tachyonic, signaling the formation of a string condensate.  In that case, the free theory does not make sense, but stabilizing interactions, such as a four-point term in the field theory, must be added.  Then the scalar fields pick up vacuum expectation values, and their masses are physical.  It turns out that our scalar field theory has no continuous global symmetry, so there are no Goldstone bosons.

The factor $1/s$ in the ``proper-time" integral corrects for overcounting strings with different starting points on the closed curves.  For unoriented strings, this factor is $1/(2s)$.  The sum over $N$ gives the usual free-field scalar partition function for oriented strings:
\begin{equation}
\label{scalarpart}
Z_{field}=\exp {\big \{}Tr\ln [\Delta_M]{\big \}}
\end{equation} 

To represent the particulars of the gauge solitons of sec.~\ref{gaugeth} we introduce $N$ scalar fields $\phi_i$, $i=1,2,\dots N$.  There are relations among these fields, following from the properties of the $Q$ matrices.  The first relation is conjugation:
\begin{equation}
\label{conj}
\phi_i^{\dagger}=\phi_{N-i}.
\end{equation}
For $N$ even, the field $\phi_{N/2}$ is self-conjugate.  The other relation is that one of the fields, say $\phi_N$, can be expressed in terms of the others, which corresponds to the vanishing of the sum of the $N$ $Q$ matrices.  In the scalar field language, this amounts to writing a vertex in the action of the form
\begin{equation}
\label{allphi}
G\int d^3x \prod_{i=1}^N\phi_i
\end{equation}
that corresponds to the vertices in Fig. (\ref{baryon}).  

It is possible in principle to calculate the properties of  the scalar-field action directly from the gauge theory.  For example, the contribution of one unit-flux vortex soliton, as displayed in Eq. (\ref{unitcv}), to the free scalar-field action $I_0$ is found by calculating the gauge action of Eq. (\ref{massterm}) based on the solitonic potential of Eq. (\ref{unitcv}).  We record here only the $B_i^2$ part; the mass term gives a similar term.
\begin{equation}
\label{unitcva}
I_0=\frac{M^3\pi}{2g^2}Tr (Q^2)\int d\tau \int d\tau '\dot{z}_i(\tau )\dot{z}_i(\tau ')\exp\{-M|z(\tau )-z(\tau ')|\}.
\end{equation}
(We do not need to indicate an index on $Q$; all $Q$s give the same result.)
The exponential term comes from the integral over all space of the square of the magnetic field strength $B_i$, which is [up to constants already absorbed in Eq. (\ref{unitcva})]
\begin{equation}
\label{bfield}
B_i(x)=\oint dz_iM^2\Delta_M(z-x).
\end{equation} 

In the infrared regime, where all length scales are large compared to $M^{-1}$, one can make contact with canonical field theory by making the large-$M$ approximations
\begin{equation}
\label{approxa}
\dot{z}_i(\tau )\dot{z}_i(\tau ')\simeq \dot{z}_i(\tau )^2,\;  |z(\tau )-z(\tau ')|\simeq |\dot{z}(\tau )||\tau -\tau '|.
\end{equation}
The integral over $\tau '$ can now be done, and (omitting exponentially small terms) results in
\begin{equation}
\label{approxa2}
I_0=Tr(Q^2)\frac{\pi M^2}{g^2}\int d\tau e^{-1}(\tau )\dot{z}_i(\tau )^2
\end{equation}
where the $einbein$ $e$ is just $|\dot{z}(\tau )|$.  This contribution to the free action is reparametrization invariant;  the form quoted in Eq. (\ref{proptime}) is the special case $|\dot{z}(\tau )|=1$.  One can add to this kinetic action a term representing energy or entropy per unit length, of the reparametrization-invariant form $\int d\tau e(\tau )M_2/2$.  By comparison of Eq. (\ref{proptime}) and Eq. (\ref{approxa2}) one sees that the mass $M_1$ is $2\pi Tr (Q^2) M^2/g^2$.  As is well-known, in $d=3$ all dynamical masses are linear in $g^2$, and therefore so is $M_1$.  

One must be careful to understand that the large-$M$ or infrared form of the action in Eq. (\ref{approxa}) is not appropriate to study the ultraviolet behavior of the scalar theory.  The contribution of one closed vortex, of whatever shape or length, to the free scalar propagator $\Delta_{0\phi}$ is the integral of $\exp [-I_0-(M_2s/2)]$ over all curves and lengths:
\begin{equation}
\label{propint}
\Delta_{0\phi}(x)=\int_0^{\infty}dse^{-M_2s/2}\int_0^x(dz)\exp{-\Big \{} \frac{M_1}{2}\int_0^s d\tau \int_0^s d\tau '\dot{z}_i(\tau )\dot{z}_i(\tau ')\exp [-M|z(\tau )-z(\tau ')|]{\Big \}} 
\end{equation}
before any approximations are made.  By writing the  integrals over $\tau$ and $\tau '$in the exponent as the limit of a sum of terms one can check that no terms singular at $s=0$ arise from the path integrals.  Normally, the path integrals lead to a factor $s^{-3/2}$ in the $s$ integral, which is responsible for ultraviolet divergences in the propagator.  But no such terms arise in our case, because the scalar field theory is built on solitons, rather than on pointlike fields.  The scalar propagator behaves at least as well as $M^2k^{-4}$ for large $k$, which means that even the $N$-point vertices discussed below lead to convergent Feynman graphs.  

The non-local (in proper time) corrections following from Eq. (\ref{propint}) to the standard local propagator of Eq. (\ref{proptime}) can be expressed as a power series in the deviation of the trajectory $z(\tau )$ from a straight line.  These are not easy to characterize, because even for a simple circular orbit the non-local terms lead to a divergent series that is not Borel-summable.  If we specify the orbit as part of a circle
\begin{equation}
\label{zorbit}
z(\tau )= R[\cos (\frac{\tau }{R}),\;\sin (\frac{\tau }{R}),\;0]
\end{equation}
the double proper-time integral becomes
\begin{equation}
\label{borel}
\int d\tau \int d\tau '\cos (\frac{\tau - \tau '}{R})\exp {\Big \{}-2MR\sin [\frac{|\tau - \tau '|}{2R}]{\Big \} }.
\end{equation}
The expansion of this integral in powers of $1/(MR)$ is a factorially-divergent series with all terms of one sign.  This expansion is closely related to the asymptotic expansion of the Bessel function $I_0(2MR)$.  We will not explore this subject further in this paper.

We next discuss the ``baryonic" vertex of Fig. (\ref{baryon}) and Eq. (\ref{allphi}).  Let us calculate the contribution $I_B$ to the logarithm of the scalar partition function $Z_{\phi}$ coming from the solitonic configuration of Fig. \ref{baryon}) [see Eq. (\ref{nsoliton}], in the infrared or local approximation.  We assume the strings in Fig. (\ref{baryon}) are well-separated on the scale of $M^{-1}$, except for the two regions of size $M^{-1}$ where they all meet.  Using the same techniques as for the propagator of a single scalar field we find   
\begin{equation}
\label{nsolitonpart}
\ln Z_{\phi}=e^{-I_N}\int d^3x \int d^3y [\Delta_{0\phi} (x-y)]^N + \dots
\end{equation}
where $I_N$ is the contribution to the action coming from the regions of size $\sim M^{-1}$ near the points $x,y$ where all $N$ strings come together.  The factor $\exp [-I_N]$ is a factor of $G^2$, where $G$ is the $N$-coupling constant in Eq. (\ref{allphi}).  

Evidently the Feynman graph represented by Eq. (\ref{nsolitonpart}) is exactly the same as the graph of the solitonic strings in Fig. (\ref{baryon}).
And this is true generally:  {\em Every graph of solitonic strings has a counterpart Feynman graph in the scalar field theory.  The sum of all connected Feynman graphs represents the logarithm of the solitonic-string partition function.}

In particular, nexuses amount to a quadratic term in the action.  The most general quadratic term is
\begin{equation}
\label{nexusact}
I_{nex}=\int d^3x \sum \phi^{\dagger}_iV_{ij}\phi_j.
\end{equation}
The diagonal terms represent single-string energy and entropy per unit length, which is the same for all vortices, and the off-diagonal terms represent the nexus-mediated  flipping of $Q_i$ to $Q_j$.  The nexus action is the same whatever the values of $i,j$.  As a result, the matrix $V$ has one element
$m_d^2$ on the diagonal and one element $m_n^2$ for every off-diagonal entry.  Such a matrix is easily diagonalized, and has $N-1$ eigenvalues $m_d^2-m_n^2$, and one eigenvalue of $m_d^2+(N-1)m_n^2$.  The $N-1$ eigenvectors of the first eigenvalue are of the form $\sum x_i\phi_i$ with $\sum x_i=0$, and for the second eigenvalue the eigenvector is $\sum \phi_i$.  In the large-$N$ limit this eigenvector does not propagate, because its eigenvalue is of order $N$.    This is consistent with the fact that in the gauge theory there are only $N-1$ unit-flux solitons. 

Finally, it is worth noting that the complete lack of ultraviolet divergences, plus the fact that all masses are linear in the gauge coupling $g^2$, is all that is needed to prove the exact result of \cite{corn94}, which is that the full quantum effective action $\Gamma$, as a function of the zero-momentum condensate $\langle G_{ij}^2\rangle$, has a minimum corresponding to a non-zero value for this condensate. 

Before going on to the rate equation derivable from this scalar field theory, we first discuss the relation of the Wilson-loop VEV to probability densities for vortices piercing the loop.

\section{\label{willoop} Wilson loop expectation value and areal densities}

We use the the GO formulation of Greensite and Olejn\'ik \cite{go}, with some minor additions.  This formulation goes beyond the usual DGA formula for relating $\sigma (k)$ to the discrete Fourier transform of the areal densities and avoids the problem, mentioned both by \cite{deldia} and GO, that if the standard DGA formula is used to relate a string tension linear in $k$ to the vortex areal densities, these densities are singular and do not show correct large-{N} behavior.  We will find a modified approach to the DGA based on GO's work which avoids this difficulty.

\subsection{\label{areal} Areal densities and the Wilson loop VEV}

Consider a large planar Wilson loop of area $A$, divided  into many squares whose size is given by the vortex correlation length $\lambda$.  We expect $\lambda \simeq 1/M$.  Let $p(L)$ be the probability that a vortex of flux $L$ exists on any square, so that the areal density of vortices is $p(L)/\lambda^2$.  We assume that only one vortex can sit on any one square, since (as discussed in the next section) two vortices attempting to occupy the same square will either repel each other or attract each other to form a new single composite vortex.  On the average there are as many antivortices as vortices, and so the antivortex probability $p(-L)\equiv p(N-L)$  is precisely equal to $p(L)$.  (So on the average the net flux through any Wilson loop is zero.  The area law of confinement arises from fluctuations in the total vortex flux.)   

The Wilson loop contains $B=A/\lambda^2$ squares, and we calculate the VEV in terms of the probabilities in the $k$ representation as
\begin{eqnarray}
\label{wvev}
\langle D_k^{-1}Tr_k\exp [\oint dx_iA_i]\rangle & = & p(0)^B+Bp(0)^{B-1}2p(1)\cos (\frac{2\pi k}{N})+\\ \nonumber
+\frac{B(B-1)}{2}p(0)^{B-2}[2p(1)\cos (\frac{2\pi k}{N})]^2 & + &
Bp(0)^{B-1}2p(2)\cos (\frac{2\cdot 2\pi k}{N})+\dots
\end{eqnarray}
where $D_k$ is the dimension of the representation $k$, and we have used the equality of vortex and antivortex probabilities.  The presence of the term $B(B-1)/2$ rather than $B^2$ multiplying $p(1)^2$ says that if two vortices are present they must occupy different sites, for reasons given above.

The sum in Eq. (\ref{wvev}) is 
\begin{equation}
\label{wvevsum}
{\big \{}\sum_{L=0}^{N-1}p(L)\exp [\frac{2\pi ikL}{N}]{\big \}}^B={\big \{}p(0)+2\sum_{L=1} 'p(L)\cos (\frac{2\pi kL}{N}){\big \}}^B
\end{equation}
where the prime on the sum indicates that it goes from 1 to $(N-1)/2$ if $N$ is odd.  If $N$ is even, the sum goes to $(N/2)-1$ and one must also add $p(N/2)\cos (\pi k)$ to the sum. 
This sum is supposed to equal the area-law result $\exp [-\sigma (k)A]$, where $\sigma (k)$ is the string tension in the $k$ representation.  Since $B=A/\lambda^2$ this leads to, as in GO and Eq. (\ref{firstgo}) 
\begin{equation}
\label{goeq}
\exp [-\sigma(k)\lambda^2]=p(0)+2\sum_{L=1} 'p(L)\cos (\frac{2\pi kL}{N})=\sum_0^{N-1}p(L)\exp [\frac{2\pi ikL}{N}].
\end{equation}
The behavior of $\sigma (k)$ is constrained by conjugation symmetry [$p(L)=p(N-L)\equiv p(-L)$].  For $k$ restricted to integer values we find
\begin{equation}
\label{sigconj}
\sigma (k)=\sigma (-k)=\sigma (N-k).
\end{equation}
This furnishes an extension of $\sigma (k)$ to negative integers.
A simple change of variables, using conjugation symmetry, shows that Eq. (\ref{goeq}) can be written as
\begin{equation}
\label{goeq2}
\exp [-\sigma(k)\lambda^2]=\sum_{-N/2}^{N/2}p(L)\cos [\frac{2\pi kL}{N}].
\end{equation}
[Here the limits $\pm N/2$ are nominal but apply for large $N$; the real limits can be worked out from the definition of the primed sum in Eq. (\ref{wvevsum}).]  

\subsection{\label{largensec} The large-N limit and conjugation symmetry}

We will be interested in the large-$N$ limit, which has certain subtleties that we explore next.

The inverse transform corresponding to Eq. (\ref{goeq} is [Eq.~(\ref{firstscale})]:
\begin{equation}
\label{invtransform}
p(L)=\frac{1}{N}\sum_{k=0}^{N-1}\exp [\frac{-2 \pi ikL}{N}]\exp [-\sigma(k)\lambda^2].
\end{equation}
 As long as $L$ is an integer it is straightforward, using conjugation symmetry as in Eq.~(\ref{sigconj}), to show that the inverse transform can also be written
\begin{equation}
\label{invtrans}
p(L)=\frac{1}{N}\sum_{-N/2}^{N/2}\exp [-\sigma(k)\lambda^2]\cos [\frac{2\pi kL}{N}].
\end{equation}
[As before, the limits on the sum are nominal.] 
From either of these equations it is very reasonable to expect that the $p(L)$ have the large-$N$  scaling property [already noted in the Introduction]
\begin{equation}
\label{scale}
p(L,N)=\frac{1}{N}\tilde{p}(\frac{L}{N})\rightarrow \frac{1}{N}\tilde{p}(x),\;x=\frac{L}{N}
\end{equation}
with conjugation symmetry implying $\tilde{p}(x)=\tilde{p}(1-x)=\tilde{p}(-x)$. 
If so, then it is a good approximation to replace the sum in Eq. (\ref{goeq}) by an integral:
\begin{equation}
\label{goeqint}
\exp [-\sigma(k)\lambda^2]=\int_{0}^{1}dx\tilde{p}(x)e^{2\pi ikx}
\end{equation}
 The error made is at least as small as $O(1/N)$, and in certain cases of physical interest is exponentially-small in $N$.   

Note that we have lost conjugation symmetry for $\sigma (k)$, but the property $\sigma (k)=\sigma (-k)$ which follows from Eq. (\ref{goeqint}) and conjugation symmetry for $\tilde{p}(x)$ replaces it for all practical purposes.   As long as $k$ is a positive or negative integer, conjugation symmetry for $\tilde{p}(x)$ allows us to write this equation in the alternative form:
\begin{equation}
\label{palt}
\exp [-\sigma(k)\lambda^2] =\int_{-1/2}^{1/2}dx\tilde{p}(x)\cos [2\pi kx].
\end{equation}

At this point we will use Eq.~(\ref{palt}) to {\em define} $\sigma (k)$ for general non-integral $k$, when we need to do so.   
  
So far we have replaced one of the discrete Fourier transforms, Eq. (\ref{goeq2}), by an integral over a finite domain, Eq.~(\ref{palt}).  This means, of course, that the inverse Fourier transform, Eq. (\ref{invtransform}), remains a sum over integral values of $k$, but the sum extends to $k=\pm \infty$:
\begin{equation}
\label{infsum}
\tilde{p}(x)=\sum_{k=-\infty}^{\infty}\exp [-2 \pi ikx]\exp [-\sigma(k)\lambda^2].
\end{equation}

One might now ask whether it is legitimate to replace the sum  in Eq. (\ref{invtrans}) by an integral with limits $\pm \infty$, provided that the sum over $k$ is well-behaved. This integral form would explicitly exhibit scaling symmetry.  But Eq.~(\ref{palt}) and this integral would  not, in general, form a Fourier transform pair.  This means that for a given $\sigma (k)$ the result of replacing the sum in Eq. (\ref{invtrans}) by an integral defines not, as one might hope, $\tilde{p}(x)$, but another function $\hat{p}(x)$, which we hope is closely related:
\begin{equation}
\label{invtrans2}
\hat{p}(x)=\int_{-\infty}^{\infty}dke^{-\sigma (k)\lambda^2}e^{-2\pi ikx}.
\end{equation}
  The general principles of Fourier integrals tell us that the true Fourier inverse that expresses $\exp [-\sigma (k)\lambda^2] $ is found from replacing the limits $\pm 1/2$ in Eq. (\ref{palt}) by $\pm \infty$ and $\tilde{p}(x)$ by $\hat{p}(x)$:
\begin{equation}
\label{fint}
\exp [-\sigma(k)\lambda^2]=\int_{-\infty}^{\infty}dx\hat{p}(x)\cos [2\pi kx].
\end{equation}
Now both $\sigma  (k)$ and $\hat{p}(x)$ are defined for certain complex domains of their arguments, and both are even functions.  The original conjugation symmetry for both of these functions has been lost; there is no reason, for example, to believe that $\hat{p}(x)=\hat{p}(1-x)$.   But it is manifest that $\hat{p}(x)=\hat{p}(-x)$.

Is there much of a relation between $\tilde{p}(x)$ and $\hat{p}(x)$?  The answer is that there can be, if in Eq.~(\ref{palt}) the integration over $x$ receives dominant contributions only from $|x|\ll$1/2, in which case $\pm 1/2$ is almost as good as $\pm \infty$ [the new limits in Eq.~(\ref{fint}).  As we will now see, the explicit formulas of GO furnish an illustration of this point, in which  smallness of the parameter $\epsilon$ introduced in Eq.~(\ref{1stepsilon})  is equivalent to effective domination of the integrals in Eqs.~(\ref{palt},\ref{fint}) by small values of $x$.  We will also see that smallness of $\epsilon$ is a form of the DGA.

The explicit phenomenological example of GO is 
\begin{equation}
\label{gotens}
\sigma (k) = k\sigma (1),\;k<\frac{N}{2};\;=(N-k)\sigma (1),\;k>\frac{N}{2}.
\end{equation} 
They then calculate the corresponding $p(L)$, which has terms exponentially-small at large $N$.  Such terms are automatically dropped  by using the discrete Fourier transform of Eq. (\ref{infsum}) with infinite limits, and we find
\begin{equation}
\label{goinfsum}
\tilde{p}(x)=\frac{1-\gamma ^2}{(1-\gamma )^2+2\gamma [1-\cos (2\pi kx)]}
\end{equation}
where
\begin{equation}
\label{gamma}
\gamma =e^{-\sigma (1)\lambda^2}=e^{-2\pi \epsilon}.
\end{equation}

Consider now what happens to GO's $\tilde{p}(x)$ [Eq.~(\ref{goinfsum})] when both $\epsilon$ and $|x|$ are small.  The result we will identify with $\hat{p}(x)$:
\begin{equation}
\label{ptilde}
\tilde{p}(x)\rightarrow\frac{\epsilon}{\pi (\epsilon^2+x^2)}\equiv \hat{p}(x).
\end{equation}
The probability $p(L)$ corresponding to $\hat{p}$ above is
\begin{equation}
\label{ptilde1}
p(L)=\frac{N\epsilon}{\pi (N^2\epsilon^2+L^2)}.
\end{equation}
This $\hat{p}(x)$, used in the Fourier integral of Eq. (\ref{fint}), yields a string tension
\begin{equation}
\label{fintinv}
\exp [-\sigma (k)\lambda^2]=\exp [-2\pi \epsilon |k|]
\end{equation}
which is the   GO input string tension  [Eq. (\ref{gotens})], provided that $k\leq N/2$.  And if the string tension is defined by Eq.~(\ref{fintinv}) for all $k$, the Fourier integral of Eq.~(\ref{fint}) yields, as it must, $\hat{p}(x)$.  The importance of small $\epsilon$ is that the poles of the integrand of Eq.~(\ref{fint}) are at $x=\pm i\epsilon$ and thus far from $\pm1/2$.

\subsection{\label{dgasec} The DGA}

Let us close this section by relating the smallness of $\epsilon$ to the DGA.  GO have observed that scaling makes it very difficult to understand how the DGA could be applicable, since $p(0)$, the probability of no vortex in a $\lambda$ square, is of order $1/N$ and so there must be vortices almost everywhere.  In the formal large-$N$ sense this is certainly true, but for fixed $N$, however large, it is always possible to choose $\epsilon \geq 1/(\pi N)$ so that $p(0)$ from Eq. (\ref{ptilde}) is   close to unity while the next-largest probability, which is $p(1)$, is an order of magnitude smaller. [One might also note that the formal $\epsilon \rightarrow 0$ limit of $\tilde{p}(x)$ in Eq. (\ref{ptilde}) is $\delta (x)$.]  Of course, in the real world of physics we cannot choose $\epsilon$, and in any case we would not expect the real world to be describable in DGA terms.  Nevertheless, it is possible to envisage a DGA that both accommodates scaling and allows for the areal densities of vortices to be fairly small.

The usual form of the DGA comes from subtracting unity from each side of the basic GO result in Eq. (\ref{goeq}), and using the normalization of probabilities to eliminate $p(0)$:
\begin{equation}
\label{dgaeq}
1-\exp [-\sigma(k)\lambda^2]=2\sum_{L=1}'p(L)[1-\cos (\frac{2\pi kL}{N})].
\end{equation} 
At least formally, the DGA follows from assuming $\sigma(k)\lambda^2$ is small and from saving only the linear term in this quantity on the left-hand side of Eq. (\ref{dgaeq}).  Since in our example $\sigma(k)\lambda^2 =2\pi \epsilon |k|$, this would require $2\pi \epsilon$ (and $2\pi k\epsilon$) to be small.   As mentioned in the introduction, ref. \cite{deldia} uses this conventional DGA to study what probabilities $p(L)$ are needed to reproduce various forms of $\sigma (k)$.  If $\sigma (k)\sim k$ for $N\gg k$, Ref. \cite{deldia} finds [in effect by linearizing the left-hand side of Eq. (\ref{dgaeq})] that for $L\geq 1$
\begin{equation}
\label{deldia}
p(L)\sim \frac{N}{L^2}.
\end{equation} 
But, as \cite{deldia} and \cite{go} note,  it is hard to understand this formula for general $L\geq 1$, because an areal density  proportional to $N$ at small ({\it i. e.}, non-scaling) values of $L$ would indicate a pileup of many vortices on a single square. The simple change from the denominator $L^2$ of Eq. (\ref{deldia}) to $L^2+N^2\epsilon^2$ in our Eq. (\ref{ptilde1}) yields a well-behaved scaling form for $p(L)$ at all $L$, including 0, while maintaining the behavior $L^{-2}$ in the regime $\epsilon N\ll L\ll N/2$. 

The basic point for us is that the simple form of $p(L)$ in Eq. (\ref{ptilde}) (1) scales as it should; (2) leads to $p(0)\sim 1/N$; and (3) has an asymptotic behavior $L^{-2}$ which is compatible with a $k$-string tension rising linearly in $k$ for small enough $k$.  This form was discovered by replacing the limit $N$ or $N/2$ on certain flux sums by $\infty$, and it is just this replacement we will make below in our approximate rate equation.  It will then be seen that this rate equation has a solution with all three properties listed above.  This is not to say that the solution of the rate equation is exactly as shown in Eq.~(\ref{ptilde}), as discussed later on.  But  our present limited investigation of the rate equation reveals no obvious imcompatibilities with Eq.~(\ref{ptilde}).

\section{\label{mastersec} The master equation}

\begin{figure}
\includegraphics[height=2.8in]{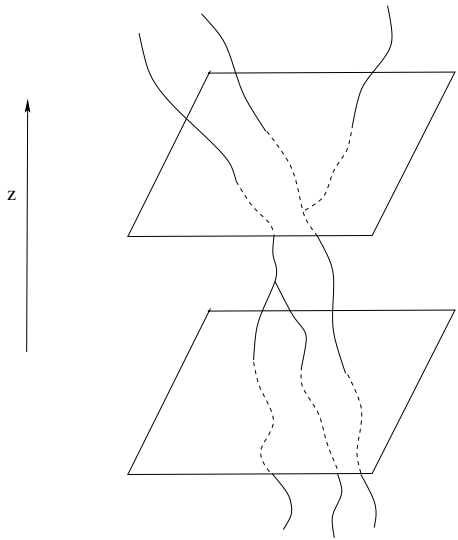}
\caption{\label{mergepict}  A test plane perpendicular to the $z$ axis is moved along that axis, revealing a continuous picture of merging and dissociating vortices piercing that plane.  Shown are two vortices recombining and another vortex dissociating as seen with $z$ as the ``time".}
\end{figure}

The master equation for vortex dynamics describes the evolution of vortex areal densities  as a function of a variable $z$, which can be thought of as giving the intercept of a test plane (essentially a planar Wilson loop), perpendicular to the $z$ axis, with that axis.  For a given configuration of branched vortices in three-space, as $z$ varies different cross-sections of the configuration are seen (see Fig. \ref{mergepict}).  These cross-sections look like a gas of $d=2$ molecules merging, dissociating, and recombining; the molecules are made of atoms that are the cross-sections of unit-flux vortices, and they attract and repel according to laws given below.  All such processes strictly conserve flux number $L\;mod\;N$.  A flux matrix $F_{\{L\}}$ is a sum of distinct $Q_i$ matrices [(see Eq. (\ref{fluxmatrix})], so we may think of any vortex of flux $L$ as made of  constituents, which we call atoms, that have unit flux and are labeled by the indices on the $Q_i$.  A center vortex of flux $L\neq 1$ is a molecule, made of those atoms which appear in the flux matrix $F_\{L\}$.  We can always take flux values to lie in the range $1\leq L \leq N-1$, and vortices with $L>N/2$ are antivortices.   On the average, as many vortices of flux $L$ as of flux $N-L$ pierce any sufficiently large plane, so the net flux is equivalent to zero.  We assume ergodic behavior, so that averaging over $z$ is the same as averaging over many different vortex-nexus configurations.   

The atoms' and molecules' motion is essentially diffusive, because they are cross-sections of random walks.  Every center vortex, of whatever flux, has the same size, roughly equal to the correlation length $\lambda$ of the vortices, so we can imagine the test plane divided into squares of area $\lambda^2$.  Each square can hold one vortex of any flux $L$, with probability $p(L)$  as described above.  The probability $p(0)$ is the probability that there is no vortex at all in a square.   

 We are interested in large $N$, first because in any case $N$ must be greater than three or there is no $k$-string problem, and second because we would like to be able to enter the scaling regime, defined in Eq. (\ref{scale}), that contains vortices whose flux $L$ scales with $N$ as $N$ grows.   As we have already mentioned, even with scaling it is possible to adjust parameters so that the probability of no vortex, $p(0)$, is considerably greater than the probability $p(1)$ of one vortex.  The scaling regime is separated from this regime of sharply-varying $p(L)$ at a flux $L\simeq \epsilon N$, where $\epsilon \ll 1$ is $\sigma (1)\lambda^2/(2\pi )$, as before. The discussion in the previous section tells us that replacing flux sums by integrals and taking certain limits of these integrals to be infinite is qualitatively appropriate, provided that $\epsilon$ is fairly small.

We will use the flux dependence of the semiclassical actions of various vortices to decide whether processes involving an action change is allowed or not.  
That we can use the semiclassical actions of center vortices at large $N$ is certainly an assumption which can be challenged, on the grounds that the actions in question diverge as $N$ gets large [see Eqs. (\ref{unitcva}), (\ref{master2})].  But one knows \cite{corn57} that the group entropy for center vortices can completely cancel the action at {\em leading} order in $N$, and that if this happens, the leading large-$N$ corrections preserve the functional form of the action except that this action is multiplied by  $const/N^2$.  The resulting then scales at large $N$.  However, the fate of non-leading large-$N$ corrections has not been studied, and we have nothing to say about them.  In any case, we must and do {\em assume} that some such phenomenon of cancellation of action and entropy takes place, or else center vortices do not survive the large-$N$ limit at all \cite{gmn}.

In any case, we make the assumption that the probabilities for particular merger and dissociation processes are large or small according to the semiclassical action difference associated with the initial and final states of merger or dissociation.   One might expect, as in the chemical law of mass action for rates, that rates depend on $\exp [-\Delta I]$ where $\Delta I$ is a difference of actions.  The rationale for this can be seen in, for example, Eq. (\ref{nsolitonpart}), where the ``baryon" vertex coupling $G$ is of the form $\exp [-I]$.  But the law of mass action strictly holds only for dilute systems, and  we take a different approach. In fact, we will drastically simplify  the rate equation by allowing flux-conserving processes for which the action difference is favorable, and forbidding those for which it is not. The actions involved depend on quantities such as $Tr(F_{\{J\}}F_{\{K\}})$, and comparing these actions for initial and final states of vortex processes yields a set of rules for probabilities of these processes, plus force laws which predict whether particular vortices will attract or repel each other.   Next we summarize allowed and forbidden processes in a set of rules for action and for forces between vortices.

\subsection{Force and action rules}

A matter of notation:  From now on, when we refer to an action, we mean (unless the context makes clear otherwise) a trace quadratic in flux matrices, without factors of $4\pi M/g^2$, etc.  So we define
\begin{equation}
\label{actiondef}
I_{\{L\}}\equiv TrF_{\{L\}}^2.
\end{equation}

Given the assumption that it is meaningful to compare the semiclassical actions, and the traces (see Appendix B) in question, we will now see that center vortices of flux $L$ can be understood as a composite of $L$ unit-flux vortices which have certain laws of attraction and repulsion.  Generally, a vortex of flux matrix $Q_i$ is repelled by another vortex of the same type, and attracted to vortices of different type but by an attraction smaller by a factor of $1/N$.  (This is to be expected for some sort of rough force balance, since there are roughly $N$ times as many ways for a vortex to see another of different type as ways to see one of its own type.)

In merger processes it is possible to form flux matrices where some $Q_i$ appear more than once.   If some indices are repeated, we say that there is an {\em overlap}.  As in Appendix B and Eq. (\ref{fluxmatrix}), we use the notation $F_{\{L\}}$ for the sum of $L$ matrices $Q_i$ with no indices  repeated.  If $R$ of the indices are each repeated once and only once, we indicate this by $F_{\{L;R\}}$.  Of course, this is an incomplete notation, since in general the precise content of the set $\{R\}$ of repeated indices must be specified.  Furthermore, one could imagine that some indices are repeated once, others repeated twice, etc.  We restrict ourselves to single repeats because the more $Q$ matrices of like type in a flux matrix, the higher the action and the less likely this is to occur. 

The actions  $I_{\{L;R\}}$ calculated as a trace in Appendix A, Eq.~(\ref{mastover}), are exactly reproduced by the following rules for calculating the self and mutual actions of unit-flux constituents:  
\begin{enumerate}
\item Each constituent (unit-flux vortex) has self-action $(N-1)/N$.
\item Unlike constituents have action $-2/N$ per interacting pair.
\item Like constituents have action +2 per pair.
\end{enumerate}
Such rules of attraction and repulsion automatically lead to a stable system, since every action in the system is the trace of a real matrix squared.

As an example, the action $I_{\{L\}}$  of a vortex of flux $L$ is $L(N-L)/N$ if there is no overlap.  We interpret it as follows.  Each of the  
$L$ constituents has self-action $(N-1)/N$, and is attracted to unlike constituents with an action contribution $-2/N$.  As a result, the action of a vortex with no overlap is
\begin{equation}
\label{actionsum}
I_{\{L\}}=\frac{L(N-1)}{N}+\frac{1}{2}L(L-1)(\frac{-2}{N})=\frac{L(N-L)}{N}.
\end{equation}

If there is overlap $R$, one must add $2R$ to this action, according to Appendix B, Eq. (\ref{mastover}).  This positive addition indicates that like constituents repel.     

We can transcribe the action and action difference associated with various processes into rules for attractive and repulsive forces between vortex atoms.
  Consider the sum of two vortices in the form given in Eq. (\ref{genvortex}), and suppose that the string for vortex $a=1$ is a straight line along the $z$ axis.  The string for vortex $a=2$ is the same, except that it is displaced in the $xy$ plane by a distance $r$.  Up to factors irrelevant for the present discussion, the action of the pair is easily found [using Eq. (\ref{master})] to have the form
\begin{equation}
\label{forceaction}
I=I_{\{K_1\}}+I_{\{K_2\}}+2e^{-Mr}[-\frac{K_1K_2}{N}+R]
\end{equation}
where $R$ is the overlap between $\{K_1\}$ and $\{K_2\}$.  Evidently, if the overlap is small enough the third term on the right is attractive, but otherwise it is repulsive.  We will use this force rule and the action rules in the master equation.

\subsection{\label{mastereq} Qualitative aspects of the master equation and its solution}

As discussed above, the master equation is a rate equation in the variable $z$, where $z$ is the point on the $z$ axis where a test plane of fixed contour and area $A$, perpendicular to this axis, intersects it (see Fig. \ref{mergepict}).

In general, there are a large number of many-vortex interactions.  To keep things manageable, we concentrate on a limited number of these, for which we furnish simple go/no go rules.  They are:
\begin{enumerate}
\item Merger of two vortices into one
\item Recombination of two vortices into two other vortices
\item Decay of one vortex into two vortices
\end{enumerate}
The rules for these processes are based on the action and force rules of the previous subsection.     In implementing the rules we will replace sums by integrals and set some upper limits in these integrals to $\infty$ rather than to $N$, in much the spirit of sec.~\ref{largensec}.

The most important action or force rule is that two vortices with zero   overlap tend to attract, and that the state of lowest action is the {\em single} vortex which combines all the constituents.  Furthermore, if they have finite overlap but still attract, they will, in whatever recombination of constituents occurs, prefer to separate the overlapped constituents entirely. That is, suppose vortices with $\{J\}=\{123456\}$ and $\{K\}=\{12789\}$ merge to form the virtual vortex $\{J+K;R\}$ with $\{R\}=\{12\}$.  Lower-action states are available, so this virtual vortex will decay quickly.  The decay state that conserves constituents and has the lowest action is the sum of vortices $\{123456789\}$ and $\{12\}$.  

Consider now a merger process in the scaling regime, in which vortices with fluxes $\{K_1\}$ and $\{K_2\}$ meet at a point (that is, they occupy the same correlation-length square in the test plane).   Neither vortex has any internal overlap (otherwise it would dissociate, as said above), but when merged there is an overlap. The  probability distribution of overlap is described in Appendix B.  The average overlap of the configuration is $\langle R \rangle =K_1K_2/N$, and it is in the scaling regime if $K_1,K_2$ are.  The merged configuration $\{K_1+K_2;R\}$ has action [in the sense of Eq. (\ref{actiondef})] 
\begin{equation}
\label{mergeaction}
I_{\{K_1+K_2;R\}}=I_{\{K_1\}}+I_{\{K_2\}}+2[R-\frac{K_1K_2}{N}].
\end{equation}
On the average, the overlap repulsion just cancels the attraction between the two vortices, so the merger is action and force neutral.  However, by recombining constituents, the action can be lowered.  The general rule is that the most favorable state is a single vortex with the combined flux of the two original vortices, as one sees from Eq. (\ref{master2}).  However, if there is overlap, one must arrange it so that $Q_i$ of the same type go to different vortices.  

If the initial configuration is rearranged into $\{L\}+\{K_1+K_2-L\}$ in such a way that neither $\{L\}$ nor its partner has internal overlap, the action of this recombined configuration, once the two vortices have separated several correlation lengths, is
\begin{equation}
\label{recombaction}
I_{\{L\}}+I_{\{K_1+K_2-L\}}=I_{\{K_1\}}+I_{\{K_2\}}-\frac{2}{N}(L-K_1)(L-K_2).
\end{equation}
The go/no go rule is that if the product $(L-K_1)(L-K_2)$ multiplying $2/N$ in this equation is positive, the recombination process is allowed, otherwise disallowed.  One easily sees that the conditions favoring the recombination are
\begin{equation}
\label{recombcond}
L<min(K_1,K_2)\;{\rm or}\; L>max(K_1,K_2).
\end{equation}
In addition, for a process to be allowed we require that there be no overlap in either the $\{L\}$ vortex or the $\{K_1+K_2-L\}$ vortex, so that
\begin{equation}
\label{recombcond2}
L>\frac{K_1K_2}{N},\;K_1+K_2-L>\frac{K_1K_2}{N}.
\end{equation}
It is important to note that the constraints of Eqs. (\ref{recombcond}, \ref{recombcond2}) are conjugation-symmetric.  Conjugation symmetry requires invariance under the exchanges
\begin{equation}
\label{conjeqn}
L \rightarrow N-L,\;K_1\rightarrow N-K_1,\;K_2\rightarrow N-K_2.
\end{equation}
Clearly, the conditions of Eq. (\ref{recombcond}) are invariant under conjugation symmetry, since they derive from the conjugation-invariant combination $(L-K_1)(L-K_2)$ of Eq. (\ref{recombaction}).  And we note that the second inequality of Eq. (\ref{recombcond2}), stemming from no overlap, can be written as the conjugation of the first inequality:
\begin{equation}
\label{conjeqn2}
K_1+K_2-L>\frac{K_1K_2}{N}\rightarrow (N-L)>\frac{(N-K_1)(N-K_2)}{N}.
\end{equation}

Now we write down the master rate equation for the probabilities $p(L,z)$.  We assume that it is a good approximation at large $N$ to replace sums by integrals, and write
\begin{equation}
\label{masteq}
\dot{p}(L,z)=\int dK_1 \int dK_2G(K_1,K_2|L)p(K_1,z)p(K_2,z)-p(L,z)\int C(L,K)p(K,z)
\end{equation}
where the dot indicates differentiation with respect to $z$.  The first term on the right-hand side of this equation gives the growth rate of $L$ vortices from recombination processes as just described, and the second term reflects the loss of vortices by two-vortex interactions.  The functions $C,G$ are, in principle, to be determined from considerations of the full set of couplings in the scalar field theory of sec.~\ref{scalarfth}. 

 Although we do not know these functions, they must obey certain conditions.  There is one relation between them, following from conservation of total probability:
\begin{equation}
\label{conserv}
0=\int dL\dot{p}(L,z)=\int dK_1 \int dK_2 p(K_1,z)p(K_2,z){\Big \{}\int dL G(K_1,K_2|L)-C(K_1,K_2){\Big \}}
\end{equation}
so that
\begin{equation}
\label{conserv2}
\int dL G(K_1,K_2|L)=C(K_1,K_2).
\end{equation}
Furthermore, the essential conjugation symmetry $p(L)=p(N-L)$ will be satisfied if the kernel $G$ obeys
\begin{equation}
\label{gconj}
G(K_1,K_2|L)=G(N-K_1,N-K_2|N-L)
\end{equation}
with a similar equation for $C$.

The probabilities we seek are those satisfying the equilibrium equation expressing independence of $z$ (which we suppress):
\begin{equation}
\label{equilib}
p(L)=\gamma \int dK_1 \int dK_2 G(K_1,K_2|L)p(K_1)p(K_2),\;\gamma^{-1} = 
\int dK C(L,K)p(K).
\end{equation}

At this point we drastically simplify the equations by taking both $G$ and $C$ to be constants $G_0,C_0$ within the regime of variables allowed by the conditions of Eqs. (\ref{recombcond}), (\ref{conjeqn2}).  Because these conditions are conjugation-symmetric, the kernels continue to obey conjugation symmetry as in Eq. (\ref{gconj}).  For suppressed processes not obeying these conditions we set $C$ or $G$ to zero.  The decay function $C$ is not subject to any particular restriction, so setting $C$ to a constant means that the parameter $\gamma$ in the equilibrium equation (\ref{equilib}) is independent of the $p(K)$, by normalization of the probabilities.  The constant $C_0$ must be $O(\lambda^{-1})$, where $\lambda$ is the correlation length in $z$.  The condition which preserves probabilities, Eq. (\ref{conserv2}), then leads to $G_0=O[(N\lambda )^{-1}]$, because $G$ is integrated over a range scaling like $N$, when its variables are in the scaling regime.  This dependence of $C,G$ on $N$ means that the rate equation can be written solely in terms of the scaled probabilities $\tilde{p}(x)$ defined in Eq. (\ref{scale}), and we will do this below.

The result is the approximate master equation
\begin{equation}
\label{constmast}
p(L)=\frac{G_0}{C_0}\int_0^NdK_1\int_0^N dK_2{\Big \{}\theta [(N-L)-\frac{(N-K_1)(N-K_2)}{N}]+
  \theta (L-\frac{K_1K_2}{N}){\Big \}}\theta [(L-K_1)(L-K_2)]p(K_1)p(K_2).
\end{equation}

 The scaling form of this equation is
\begin{equation}
\label{approxmast}
\tilde{p}(x)=\tilde{p}(0)\int_0^1dx_1\int_0^1dx_2{\Big \{}\theta [1-x-(1-x_1)(1-x_2)]+ \theta (x-x_1x_2){\Big \}}\theta [(x-x_1)(x-x_2)]\tilde{p}(x_1)\tilde{p}(x_2)
\end{equation}
  where we have made the replacement
\begin{equation}
\label{gceq}
\frac{G_0}{C_0}\equiv \frac{\tilde{p}(0)}{N}.
\end{equation}
 That this is correct follows from Eq. (\ref{approxmast})  by setting $x=0$ in the rate equation, which yields Eq.~(\ref{gceq}).

 It is not our purpose in the present paper to study the rate equations in any great detail, largely because details of the solution will depend on details of the kernels $C,G$, which we do not know.  However, there are certain features of the rate equations that we believe are robust:   The behavior near zero, which is consistent with scaling, as we have already seen, and the behavior for flux $L$ which is large compared to unity but still small compared to $N/2$.  This is the regime of the modified DGA, already studied in sec.~\ref{largensec}, in which limits such as $N$ or $N/2$ were replaced by $\infty$, equivalent to replacing limits in the scaling variable $x$ of $\pm$1/2 by $\pm \infty$.  So in Eq.~(\ref{constmast}), one sets the explicit $N$s inside the integral to infinity (but not the $N$ in the factor $G_0/C_0$), and in the scaling equation (\ref{gceq}) one replaces the limit $x=1$ by $x=\infty$.   After a little algebra, in the $N\rightarrow \infty$ limit Eq.~(\ref{constmast}) becomes
\begin{equation}
\label{constmast2}
p(L)=p(0)\int_0^{L}dK_1\int_{L-K_1}^Lp(K_1)p(K_2)+p(0)\int_L^{\infty}dK_1\int_L^{\infty}dK_2p(K_1)p(K_2).
\end{equation}
[In this equation there is a factor of $N^{-1}$ hiding in $p(0)$ and therefore a factor of $N$ hiding in $p(L)$.]

 For large $L$ it is elementary to find the self-consistent behavior of the second term on the right-hand side of Eq.~(\ref{constmast2}), since in the integral only large values of $K_1,K_2$ contribute.  The first term on the right has contributions from this regime but also from the regime where one or the other of $K_1,K_2$ is small (if both are small, they do not contribute to large $L$).    In this second regime we assume that $p(K)$ is constant and of order $p(0)$ for $K\leq p(0)^{-1}$, and takes on its large-$K$ behavior for $K\sim L$.  After some analysis one finds that the self-consistent behavior at large $L$  of $p(L)$ is simply that of Eq. (\ref{deldia}) and already given in Ref. \cite{deldia}, except that we can supply an overall factor:
\begin{equation}
\label{deldia2}
p(L)= {\rm const}\frac{N}{\tilde{p}(0)L^2}.
\end{equation}
or equivalently
\begin{equation}
\label{scalesoln}
\tilde{p}(x)={\rm const}\frac{1}{\tilde{p}(0)x^2}.
\end{equation}
where the constant factor is of order unity.   One need do no analysis at all if one accepts that $p(L)$ has a power-law behavior at large $L$, since $L^{-2}$ is the only possible such behavior.

This large-$L$ behavior is quite acceptable in the scaling regime $L\sim N$, leading to densities uniformly of order $1/N$.  But there are certainly corrections for $L$ in the non-scaling regime, since as already pointed out $\tilde{p}(0)$ is finite.   The {\em simplest} extrapolation that shows the asymptotic behavior at large $L$ of Eq. (\ref{deldia2}) as well as the correct qualitative behavior at $L=0$ is the one already used phenomenologically in Eq. (\ref{ptilde}):
\begin{equation}
\label{extrap}
p(L)=\frac{\epsilon N}{\pi [(\epsilon N)^2+L^2]},\;\tilde{p}(x)=\frac{\epsilon}{\pi (\epsilon^2+x^2)}.
\end{equation}
It follows that 
\begin{equation}
\label{tildep0}
\tilde{p}(0)=\frac{1}{\pi \epsilon}
\end{equation}
and so the normalization of the large-$L$ or large-$x$ behavior in Eq.~(\ref{deldia2})  leads to the same parametric behavior in $\epsilon$ as does the extrapolation formula given in Eq.~(\ref{extrap}).  
 Of course,  we already know from  sec.~\ref{largensec}, or if one accepts the DGA from \cite{deldia}, that this extrapolation leads to a string tension linear in $k$, at least for $ k\ll N$.

It is simply not possible to argue for or against the extrapolation formula of Eq. (\ref{extrap}) without knowing more than we know about the kernels $C,G$ of the master equation.   Let us, for example, modify the master equation (\ref{approxmast}) by replacing $G_0$ by a function $G(L)$.  In principle, the idea is that we are given $G(L)$ and try to solve Eq. (\ref{approxmast}) for $p(L)$.  The solution is not obvious; it requires solving a non-linear integral equation.  But one can think of the problem the other way around:  Given $p(L)$, to what kernel $G(L)$ does it correspond?  This question is answered by doing some integrals.  If we use our guessed extrapolation in Eq. (\ref{extrap}) we find that $G(L)$ is a smooth and slowly-varying function approaching constant values as $L\rightarrow 0,\infty$, no more or less plausible than assuming it is a strict constant. 

One may wonder how to recover conjugation symmetry from a non-conjugation-symmetric result such as the extrapolation of Eq.~(\ref{extrap}).    One could make the substitution
\begin{equation}
\label{conjsub}
L^2\rightarrow \frac{N^2}{2\pi^2}[1-\cos (\frac{2\pi L}{N})].
\end{equation}
The right-hand side above is indeed $L^2$ in the regime of interest to us:  $ L \ll N/2$.  Our extrapolation equation then becomes
\begin{equation}
\label{newextrap}
p(L)=[\frac{const}{N}] \frac{4\pi^2\epsilon}{(2\pi \epsilon )^2+2-2\cos (\frac{2\pi L}{N})}.
\end{equation}
To no one's surprise, this is  precisely of the GO form given earlier in Eq.~(\ref{goinfsum})  if one identifies
\begin{equation}
\label{gocomp}
2\pi \epsilon = 2\sinh (\frac{\sigma (1)\lambda^2}{2}),
\end{equation}
which in the small-$\epsilon$ limit gives  our previous result.

\section{\label{sumsec}Summary and conclusions}

Starting from the effective $d=3$ gauge-theory action which yields center vortices and nexuses as solitons, we construct an $N$-component scalar field theory.  The components of this field theory are the unit-flux constituents  (whose $d=2$ cross-sections in a test plane we call atoms) of center vortices having higher flux $J$, whose $d=2$ cross-sections  we call molecules.  Like constituents repel, and unlike constituents attract, with an attraction $1/N$ times the repulsion.  This leads to a condensate of merging, dissociating, and recombining vortices, and an approximate qualitative picture of vortex areal densities, based on studying these processes in a test plane that is pierced by vortices.  This study leads in turn to a master rate equation, whose general structure we can foresee but whose detailed quantitative properties remain to be determined in future work.

  We expect that  the basic dynamics of the $d=2$ interaction of our atoms and molecules is the same in $d=4$ and $d=3$,  and the master rate equation will have only minor quantitative differences in these two dimensions. 
  
We also give a rough and tentative first investigation of the rate equation, based on an approach to large-$N$ dynamics which in some sense generalizes the usual implementation of the DGA.  This generalization makes use of the work of GO, who propose a relation between $k$-string tensions and areal densities (or equivalently probabilities) that is not at all restricted by the DGA.  Its utility is that we can to a large extent bypass the complications of conjugation symmetry.  At large $N$ we replace discrete Fourier transforms by integrals with infinite limits, and show that the validity of this substitution defines a modified form of the  DGA.  These integrals preserve the scaling property $p(L)=(1/N)\tilde{p}(L/N)$.  Conjugation symmetry [$p(L)=p(N-L)$, etc] is violated, but replaced by an equivalent condition that requires both $p(L)$ and $\sigma (k)$ to be even functions of their argument.  We argue that this replacement of sums by infinite integrals violates no qualitative feature of the $k$-string tension problem, provided that the scaling argument $L/N$ is small compared to unity, and that $k\ll N/2$.

The rate equation, using some simplified kernels, has  a solution $p(L)$  that obeys scaling; which therefore has a probability of no vortex $p(0)$ that is $O(1/N)$; and has an asymptotic behavior $NL^{-2}$ (as proposed earlier in \cite{deldia} on phenomenlogical grounds). 
The simplest extrapolation of our asymptotic results to all fluxes is precisely the form taken by the GO phenomenological form when $L$ is large but obeys $L\ll N/2$, and when a parameter $\epsilon$ defining the modified DGA is small.  This extrapolation   is such that the $k$-string tension is linear in $k$ for  $k\ll N$.  In this paper we leave entirely open the closely-related questions of the behavior  which restores conjugation symmetry (requiring understanding the behavior for $k>N/2$), and of non-leading corrections to the large-$N$ limit.

Work is in progress to refine the drastic simplifications of the present paper, by calculating (from the gauge-theory effective action) those solitonic actions necessary to compute the scalar field theory coupling constants which describe merging, dissociating, and recombining vortices.  Then one can hope to write down a more accurate master rate equation, properly including such effects as conjugation symmetry, and solve it numerically. 

Even without the elaborate work of calculating all the solitonic actions, it would be valuable to carry out computer simulations of the dynamics of atoms and molecules subject to the simple rules of attraction and repulsion used in this paper,   incorporating all $n$-body processes of merger, dissociation, and recombination as well as conjugation symmetry and $mod\;N$ flux conservation.  

It would also be of great value to study lattice simulations for moderately-large values of $N$, say, $N=4,5,6$, and look for center vortices of various fluxes undergoing merging, dissociating, and recombining.  In this way one could make a connection between direct simulations of $k$-string tensions and the behavior of the vortex condensate.

\appendix

\section{Properties of vortex flux matrices}

We introduce the $N\times N$ traceless diagonal matrices
\begin{equation}
\label{qmat}
Q_i={\rm diag}(\frac{1}{N},\;\frac{1}{N}, \dots \frac{1}{N},\;-1+\frac{1}{N},\;\frac{1}{N},\;\dots)\;\;i=1,2,\dots N
\end{equation}
with the -1 in the $i$th position.   Although $N$ matrices are defined, their sum is zero, as is easily checked:
\begin{equation}
\label{qsum}
\sum_1^N Q_i = 0.
\end{equation}
The independent $Q_i$ are just linear combinations of the weight vectors $\vec{\mu}_i$ for the fundamental representation of $SU(N)$.  The $Q_i$ obey
\begin{equation}
\label{qnorm}
Tr (Q_iQ_j)=\delta_{ij}-\frac{1}{N}
\end{equation}
which is twice the scalar product $\vec{\mu}_i\cdot \vec{\mu}_j$.

Of course, any $Q_i$ is related to any other by a similarity transformation $Q_i=S[ij]Q_jS[ij]^{-1}$.  The $S[ij]$ and their products form a representation of the permutation group, which means that various group orthogonality properties hold in summing over the vortex collective coordinates represented by the permutation group.

The flux associated with each $Q_i$ is $2\pi /N$, in the sense that
\begin{equation}
\label{qexp}
\exp [2\pi i Q_j]=\exp [2\pi i/N] 
\end{equation}
for all $j$. Equation (\ref{qsum}) simply means that flux is conserved $mod\;N$.  

Next we define flux matrices for higher flux via
\begin{equation}
\label{feq}
F_{\{J\}}\equiv Q_{i_1}+\dots Q_{i_J}
\end{equation}
where no two indices are the same. If $R$ of the indices occur at least twice each, we use the notation $F_{\{J;R\}}$. Of course, to specify the detailed structure of the overlap we must write  the index set in Eq. (\ref{feq}) explicitly:  $\{J\}=\{i_1\dots i_J\}$.  For example, the set $\{J\}=\{111223334567\}$ has an overlap $R=3$, but simply specifying $R$ does not specify the exact overlap.  In practice (see below) vortices with a given flux $J$ and non-zero overlap always have higher action than for the same flux and no overlap, so in this paper {\em we will restrict overlapped index sets to those in which an index in the overlap set $\{R\}$ occurs exactly twice.}  The example above is not allowed, then, but $\{J\}=\{1122334567\}$ is; it has overlap $R=3$. Obviously, this restriction is easy to remove, but the consequent notation becomes cumbersome.  

The collective coordinates for $F_{\{J\}}$ are permutations of the matrix indices---the same permutation for each $Q_i$ in the sum.  In particular, if the set $\{J\}$ is held fixed, and one sums over all permutations of the indices in another set $\{K\}$, there is orthogonality:
\begin{equation}
\label{collcoord}
\sum_{permK}Tr[F_{\{K\}}F_{\{J\}}]=0.
\end{equation}

Note that there is a close connection between center vortices of flux $\{J\}$ in the fundamental representation of $SU(N)$ and elementary vortices (unit flux) in the $J$ (totally antisymmetric) representation of this group.  Every state of the $J$ representation can be labeled by indices running from 1 to $N$ as $|i_1\dots i_J\rangle $, with no two indices alike, and ket vectors with the same index set but in different order differ by the sign of the permutation required to go from one order to the other.  For any diagonal matrix $D={\rm diag}(D_i,\dots D_N)$ in the fundamental representation, the action of $D$ in the $J$ representation on a ket is
\begin{equation}
\label{jrep}
D|i_1\dots i_J\rangle =(D_{i_1}+\dots D_{i_J})|i_1\dots i_J\rangle.
\end{equation} 
We have the equation
\begin{equation}
\label{recip}
\frac{1}{C_JD_J}Tr_JF_{\{L\}}^2=\frac{1}{C_1D_1}Tr_1F_{\{L\}}^2
\end{equation}
where $C_J=J(N-J)(N+1)/(2N)$ is the quadratic Casimir of the $J$ representation, and $D_J=N!/[J!(N-J)!]$ is the dimension of this representation (note that $J=1$ is the fundamental representation).   On the left-hand side of Eq. (\ref{recip}) traces are in the $J$ representation; on the right-hand side, in the fundamental representation.

So far the flux labels $J$ are positive numbers.  From Eq. (\ref{qsum}) it follows that
if $\{J\}$ and $\{N-J\}$ have no overlap
\begin{equation}
\label{modn}
F_{\{J\}}+F_{\{N-J\}}=0,\;F_{\{-J\}}\equiv -F_{\{J\}} = F_{\{N-J\}}.
\end{equation}
One sees that
\begin{equation}
\label{qexp2}
\exp [2\pi iF_{\{\pm J\}}]=\exp [\pm 2\pi iJN].
\end{equation}

The master formula from which all other quadratic trace formulas follow is
\begin{equation}
\label{master}
Tr (F_{\{J\}}F_{\{K\}})=\frac{-JK}{N}+R
\end{equation}
if $\{J\}$ and $\{K\}$ have overlap (intersection) $\{R\}$.  In this equation, both $J$ and $K$ are taken as positive numbers, so that if one uses this equation for a negative flux $F_{\{-J\}}$, one replaces this by $-F{\{J\}}$. Or one comes to the same thing by using $F_{\{N-J\}}$ in place of  $-F{\{J\}}$, but then one must replace the overlap by $\bar{R}=K-R$.

The action of a center vortex $\{J\}$ with no  overlap  is found by setting $\{J\}=\{K\}$, yielding
\begin{equation}
\label{master2}
Tr F_{\{J\}}^2=\frac{J(N-J)}{N}.
\end{equation}
When there is overlap (as restricted above) we write the set $\{J;R\}$ as
the sum of two sets with no internal overlap, but of course the two sets overlap each other:  $\{J;R\}=\{J-R\}+\{R\}$.  One then finds the action for $\{J;R\}$ to be
\begin{equation}
\label{mastover}
Tr F_{\{J;R\}}^2=\frac{J(N-J)}{N}+2R.
\end{equation}

\section{Overlap probability}

Consider the temporary merger of vortices $\{J\}$ and $\{K\}$, which have no internal overlaps but which taken together have overlap $\{R\}$.  An example is $\{J\}=\{12345\}$, $\{K\}=\{45678\}$, $\{R\}=\{45\}$.  The composite vortex $\{J+K;R\}$ has higher action [Eq. (\ref{mastover})] than a composite with the same total flux but with $\{R\}=\{0\}$.  We wish to know, given $J,K$, the distribution of overlap number $R$.  Evidently, given a set of $K$ numbers taken from 1 to $N$, with no repeats allowed, the probability that one of an independent set of $J$ numbers is in the set $\{K\}$ is $K/N$.  This suggests that (just as in tossing coins with a probability $K/N$ of getting heads) the average overlap in $J$ trials is
\begin{equation}
\label{overlap}
\langle R \rangle =\frac{JK}{N}.
\end{equation}

The complete probability distribution of $R$, which we call $P(R|J,K)$, is found easily. If the overlap is indeed $R$, it occurs a number of times equal to the number of ways of picking $J$ numbers subject to the constraint that $R$ of them are in the set $\{R\}$, that is, the number of ways to choose $J-R$ out of ${N-K}$. This is to be multiplied by the number of ways an overlap of $R$ numbers can be imbedded in $\{K\}$, which is the number of ways of picking $R$ out of $K$, and divided by the total number of ways of picking $J$ out of $N$.  The result is
\begin{equation}
\label{overlap2}
P(R|J,K)=\frac{J!K!(N-J)!(N-K)!}{N!R!(N-J-K+R)!(K-R)!(J-R)!}.
\end{equation}
Using Stirling's formula, with the assumption $N\gg J,K,R\gg 1$, leads to a binomial distribution that is equivalent to a Poisson distribution:
\begin{equation}
\label{overlap3}
P(R|J,K)=\frac{J!}{R!(J-R)!}(\frac{K}{N})^R(\frac{N-K}{N})^{J-R}\rightarrow
\frac{1}{R!}(\frac{JK}{N})^Re^{-JK/N}.
\end{equation} 
Note that the probability of no overlap ($R=0$) is $\exp [-JK/N]$, which is exponentially small in the scaling regime, and that the expectation value of $R$ is indeed $JK/N$.

\end{document}